\documentclass[12pt]{article}
\pdfoutput=1
\usepackage[latin1]{inputenc}

\usepackage{amsmath}
\usepackage{amsfonts}
\usepackage{amssymb}
\usepackage{graphicx}
\usepackage{geometry}
\usepackage{amssymb,epsfig,subfigure}
\usepackage{hyperref}

\makeatletter
\renewcommand\section{\@startsection {section}{1}{\z@}%
                                 {-3.5ex \@plus -1ex \@minus -.2ex}
                                   {2.3ex \@plus.2ex}%
                                   {\normalfont\large\bfseries}}
\renewcommand\subsection{\@startsection{subsection}{2}{\z@}%
                                   {-3.25ex\@plus -1ex \@minus -.2ex}%
                                     {1.5ex \@plus .2ex}%
                                     {\normalfont\bfseries}}
\renewcommand\subsubsection{\@startsection{subsubsection}{3}{\z@}%
                                   {-3.25ex\@plus -1ex \@minus -.2ex}%
                                     {1.5ex \@plus .2ex}%
                                     {\normalfont\itshape}}
\makeatother

\def\pplogo{\vbox{\kern-\headheight\kern -29pt
\halign{##&##\hfil\cr&{\ppnumber}\cr\rule{0pt}{2.5ex}&\ppdate\cr}}}
\makeatletter
\def\ps@firstpage{\ps@empty \def\@oddhead{\hss\pplogo}%
  \let\@evenhead\@oddhead 
}
\def\maketitle{\par
 \begingroup
 \def\thefootnote{\fnsymbol{footnote}}
 \def\@makefnmark{\hbox{$^{\@thefnmark}$\hss}}
 \if@twocolumn
 \twocolumn[\@maketitle]
 \else \newpage
 \global\@topnum\z@ \@maketitle \fi\thispagestyle{firstpage}\@thanks
 \endgroup
 \setcounter{footnote}{0}
 \let\maketitle\relax
 \let\@maketitle\relax
 \gdef\@thanks{}\gdef\@author{}\gdef\@title{}\let\thanks\relax}
\makeatother

\numberwithin{equation}{section}

\newcommand{\be}{\begin{equation}}
\newcommand{\bea}{\begin{eqnarray}}
\newcommand{\ee}{\end{equation}}
\newcommand{\eea}{\end{eqnarray}}
\newcommand\beq{\begin{equation}}
\newcommand\eeq{\end{equation}}

\newcommand{\mc}{\mathcal}

\begin{document}

\setcounter{page}0
\def\ppnumber{\vbox{\baselineskip14pt
}}
\def\ppdate{\footnotesize{SLAC-PUB-14429 ~~~ SU-ITP-11/30}} \date{}

\author{Ning Bao, Xi Dong, Eva Silverstein, and Gonzalo Torroba\\
[7mm]
{\normalsize \it Stanford Institute for Theoretical Physics }\\
{\normalsize  \it Department of Physics and SLAC, }\\
{\normalsize \it Stanford, CA 94309, USA}\\
[3mm]}

\bigskip
\title{\bf  Stimulated superconductivity at strong coupling
\vskip 0.5cm}
\maketitle

\begin{abstract}
Stimulating a system with time dependent sources can enhance instabilities, thus increasing the critical temperature at which the system transitions to interesting low-temperature phases such as superconductivity or superfluidity.  After reviewing this phenomenon in non-equilibrium BCS theory (and its marginal fermi liquid generalization) we analyze the effect in holographic superconductors.  We exhibit a simple regime in which the transition temperature increases parametrically as we increase the frequency of the time-dependent source.

\end{abstract}
\bigskip
\newpage

\tableofcontents

\vskip 1cm

\section{Introduction}\label{sec:intro}

Considering time-dependent couplings among quantum fields leads to
a rich theory and phenomenology with a number of important applications.\footnote{not least among them the origin of structure in the universe.}  One interesting effect was explained early on in \cite{Eliashberg},  following the development \cite{BCS}\ of a microscopic theory of superconductivity as pairing of charged fermions leading to spontaneous breakdown of electromagnetic gauge symmetry.  The work \cite{Eliashberg}\ argued that time dependent high-frequency electromagnetic fields could modify the density of states of the fermions in such a way as to increase the transition temperature $T_c$ of the material (without appreciably heating it up), and the work \cite{ChangScalapino}\ analyzed the net effect taking into account phonon interactions.  This effect had been seen experimentally, and has been tested more recently with positive results reported \cite{Tinkham}. Recently there appeared work \cite{SF}\ analyzing the closely analogous effect in superfluids built from cold atoms, which may have the advantage of additional experimental control.  Other interesting low-temperature phases can also be stimulated in a similar manner \cite{othertdep}.  In each of these works, the calculations were controlled by treating the externally applied time-dependent sources perturbatively.

Abstractly in quantum field theory, it is easy to see how time-dependent couplings can enhance an instability such as a superconducting transition.  Consider a scalar field with a time-dependent coupling $g(t)$
\beq\label{tdep}
S=\int d^4 x \frac{1}{g(t)^2} \left\{(\partial\phi)^2 -\frac{1}{2}m^2\phi^2 +{\cal L}_{interaction} \right\}.
\eeq
Canonically normalizing the scalar field by setting $\tilde\phi\equiv\phi/g(t)$ yields
an action
\beq\label{tdepII}
S=\int d^4 x\left\{(\partial\tilde\phi)^2 - \left[m^2-2\left(\frac{\dot g}{g}\right)^2+\frac{\ddot g}{g}\right]\tilde\phi^2+{\tilde{\cal L}}_{interaction} \right\}.
\eeq
The quantity in square brackets behaves as an effective mass squared, $m_{{\rm eff}}^2$. The shift is caused by the time-dependent coupling.  For a range of functions $g(t)$, the shift is negative.  In the context of a Ginzburg-Landau model, this reduction in $m_{{\rm eff}}^2$ enhances the instability, raising the transition temperature $T_c$ below which the scalar field condenses.  (Depending on the implementation, it may also thermally excite the system, so determining the net effect requires more detailed analysis.)

In practical applications, one does not \emph{a priori} control $g(t)$ in the low-energy effective theory;  one may either control the chemical potential as a function of time, or irradiate a sample electromagnetically, or both.   We are therefore motivated to analyze the effects of such time-dependent perturbations in more general quantum field theories, in particular strongly coupled field theories which exhibit superconducting transitions~\cite{Hartnoll:2008vx}.

In \S\S 3 and 4 we will derive simple (but controlled) analytic estimates exhibiting our effect, and check them with appropriate numerical solutions.  By working in a simple probe limit and estimating the particle production and decays, we will check that for a reasonable class of examples the main effect of the time dependence is to raise the effective $T_c$ of the system (as opposed to heating it up).\footnote{Other interesting works on time-dependent finite density holographic field theory have appeared recently, including \cite{otherholtdep}.  These have typically focused on a return to equilibrium, whereas in this work we are concerned with the effect of nonequilibrium dynamics in enhancing phase transitions.} In modern pump-probe experiments, one can study non-equilibrium dynamics, but one typically heats up the system. It would be interesting to model this physics holographically as well.

Before turning to that, however, we will briefly review the effect in BCS theory and comment on its generalization to strongly coupled analogues along the lines of \cite{Faulkner:2010tq}.
\footnote{Although no complete model of high-temperature superconductivity has been derived, it is interesting
to consider the corresponding question in those materials.
Ultimately a factor of 2 in $T_c$ would make the difference between current high-$T_c$ superconductors and room-temperature ones.}

\section{Time Dependence and Pairing}\label{sec:FP}

In this section we briefly review the mechanism of \cite{Eliashberg}, originally applied to BCS pairing in Fermi liquid theory, and comment briefly on its generalization to the non-Fermi liquid model of \cite{Faulkner:2010tq}.
In BCS superconductors \cite{BCS}, one calculates the effective potential for a Cooper pair; the resulting equation of motion is the gap equation, essentially the equation of motion for the charged scalar field condensate $\Delta$, which takes the form
\beq\label{gap}
\Delta=\lambda N(0)\int^{E_{\rm cutoff}} dE\frac{\Delta}{(E^2+\Delta^2)^{1/2}}\left(1-2n_{FD}\right).
\eeq
Here $N(0)$ is the density of states at the Fermi surface ($E\to 0$), $\lambda$ the strength of the pairing interaction, and $n_{FD}=1/(e^{E/T}+1)$ the Fermi-Dirac distribution as a function of temperature $T$.  For $T$ larger than a critical temperature $T_c\sim E_{\rm cutoff}e^{-1/\lambda N(0)}$, the only solution to (\ref{gap}) is $\Delta=0$.  Below $T_c$ there is a nontrivial solution for $\Delta$, leading to the Higgs mechanism and superconductivity (see \cite{Weinberg:1986cq}\ for a pedagogical review).

The kinematics of the Fermi surface is crucial in (\ref{gap}); the effect dies away exponentially as the density of states $N(0)$ at the Fermi surface goes to zero.
The finite-temperature distribution $n_{FD}$ smooths out the Fermi surface and eventually eliminates the superconductivity.   Thermal occupation of quasiparticle states just above the original Fermi surface prevents them from contributing to (\ref{gap}).
Eliashberg \cite{Eliashberg}\ noted that time dependent perturbations could counteract this effect of finite temperature, by exciting the quasiparticles just above the original Fermi surface to significantly higher levels. This effectively shifts the distribution function $n_{FD}$ in (\ref{gap}) to a smaller value,
\beq\label{shiftn}
n_{FD}\to n_{FD}+\delta n \;\;\;{\rm with} \;\;\;\delta n<0.
\eeq
With the excited quasiparticles out of the way, the system can access the states near the original Fermi surface, enhancing pairing and condensation.  Chang and Scalapino \cite{ChangScalapino}\ and others studied the further effects of phonon-quasiparticle interactions, finding that additional enhancements are possible.

In the next section, we analyze the effects of time dependent perturbations on holographic superconductors.\footnote{More precisely, the gravity side is dual to a superfluid, but this distinction will not be important in our case. Some of the nonequilibrium effects are in fact easier to observe in superfluids.}  In these systems, we will find that time dependent perturbations also enhance $T_c$, even allowing for superconductivity when the average chemical potential vanishes.  Although it goes in the same direction, the mechanism behind this is not the same in detail as that in the BCS example.  One question raised in the earlier works (see e.g. \cite{SF}) is how to extend the calculations beyond a perturbative treatment in $\delta n$ (equivalently, perturbative in the original time dependent source) to see how large the enhancement of $T_c$ could become.  In our case, the holographic methods allow us to analyze large $\delta n$, though in an unrealistic regime of large-N gauge theory.  As we will see, the effect can be parametrically large in this regime.

An example that may be closer to the BCS case is to consider pairing and time dependent perturbations in the free sector of the model \cite{Faulkner:2010tq}, which produces a marginal Fermi liquid upon mixing with a putative locally critical strongly coupled sector of fermion operators.  Time dependent perturbations could again be used to excite thermally distributed states away from the original Fermi surface, increasing $T_c$ from its time-independent value.

\section{Holographic Time-Dependent Superconductors}\label{sec:time}

In this section, we analyze the effect of a time-dependent chemical potential on $T_c$ in a holographic superconductor of the type introduced in~\cite{Hartnoll:2008vx}.\footnote{Effects of a spatially dependent chemical potential on holographic superconductors were studied in \cite{Flauger:2010tv}.} We will take the temperature $T \approx T_c$, a limit which is sufficient for obtaining the critical temperature. This corresponds to the onset of the instability and reduces the nonlinear system to a single linear differential equation. In \S \ref{sec:condensed} we will study the condensed solution that arises when $T< T_c$.

The basic framework is 4D AdS gravity with a $U(1)$ gauge field and a charged scalar:
\be\label{eq:S1}
S= \int d^4 x \sqrt{-g} \left(R + \frac{6}{L^2}-\frac{1}{4} F_{\mu\nu}F^{\mu\nu}- |\partial \Phi - i q A \Phi|^2-m^2 |\Phi|^2 \right)\,,
\ee
where $L$ is of order the $AdS$ radius.
The background metric corresponds to an AdS-Schwarzschild black hole,
\be\label{eq:ds2}
ds^2 = -f(r) dt^2 + \frac{dr^2}{f(r)}+\frac{r^2}{L^2} (dx_1^2+dx_2^2)\;\;,\;\;f(r)=\frac{r^2}{L^2}\left( 1- \frac{r_+^3}{r^3}\right),
\ee
where the temperature is
\be\label{eq:T}
T= \frac{3r_+}{4\pi L^2}\,.
\ee

The dual QFT has an operator $\mc O$ dual to $\Phi$, and its expectation value is the order parameter for superconductivity. The dimension $\Delta$ of $\mc O$ and the mass of $\Phi$ are related by
\be\label{eq:Delta}
\Delta(\Delta-3)= m^2 L^2\,.
\ee
The $U(1)$ bulk gauge field is dual to a global symmetry current in the QFT, so strictly speaking (\ref{eq:S1}) describes a superfluid. This distinction, however, will not play an important role here, and it is possible to make the boundary gauge field dynamical. We will comment more on this below.

We will consider a time-dependent chemical potential in the
dual QFT:
\be
A_t \to \mu(t)\;,\;\textrm{for}\;\;r\to \infty,
\ee
and we want to find its effects on the charged scalar field $\Phi$.

A time-independent chemical potential has been analyzed in~\cite{Hartnoll:2008vx} and other works.  The coupling of the charged scalar $\Phi$ to the gauge potential $A_t$ leads to a negative contribution to the effective mass squared for $\Phi(r)$, causing it to condense for $r<r_c$, where $r_c$ is the radial position below which the effective mass squared is sufficiently negative.
In the present work we wish to compute the contribution of time dependent effects to the effective mass squared, and determine the new $r_c$ below which $\Phi$ condenses.  This provides a measure of the shift of $T_c$ in the dual QFT.

\subsection{Probe Limit}\label{subsec:probe}

To illustrate the effect, it is sufficient to work in the probe limit where $A_\mu$ and $\Phi$ do not backreact on the metric. This entails adding stress-energy sources in the bulk which are subdominant to $M_P^2/L^2$, where $M_P$ is the bulk four-dimensional Planck mass scale.\footnote{More generally, other scales may arise in the UV completion of the system, for which similar remarks hold.} In this regime, the equations of motion are
\bea
\nabla_\mu F^{\mu\nu}&=& iq (\bar \Phi \partial^\nu \Phi - \Phi \partial^\nu \bar \Phi)+2q^2 |\Phi|^2 A^\nu\nonumber\\
D_\mu D^\mu \Phi &=&m^2 \Phi\,.
\eea

To begin, consider an ansatz where the only nonzero gauge field is $A_t$. In the static case, the phase of $\Phi$ is required to be constant, something which is no longer true once time-dependence is allowed; for instance, Maxwell's equation along the $r$-direction now reads
\be
\partial_t \partial_r A_t = i q f(r) (\bar \Phi \partial_r \Phi - \Phi \partial_r \bar \Phi)\,.
\ee
The dynamics is then easier to analyze in unitary gauge, where a gauge transformation is used to set the phase of $\Phi$ to zero. The gauge transformation achieving this is
\be
A_\mu^{'}=A_\mu - \frac{1}{q} \partial_\mu(\arg(\Phi))\,.
\ee
Henceforth we work in terms of $A'$ and drop the primes; in this gauge, we need to allow for both $A_t$ and $A_r$ components.

In unitary gauge the equations of motion simplify further. Maxwell's equations along the time and radial directions now read
\bea\label{maxwelleom}
& &\frac{1}{r^2}\partial_r(r^2F_{rt})-\frac{2 q^2}{f} \Phi^2 A_t=0\\
\label{maxwelleom2}
& &\partial_t F_{rt} = 2 q^2 f \Phi^2 A_r\,.
\eea

The second equation shows explicitly that the source for $A_r$ is given by the time and radial dependence of $A_t$. Note that because of our unitary gauge choice, $\Phi$ is real here. Its equation of motion is
\be\label{phieom}
\frac{\partial^2 \Phi}{\partial r^2}+ \left(\frac{\partial_r f}{f}+\frac{2}{r} \right)\frac{\partial \Phi}{\partial r}-\frac{1}{f^2}\frac{\partial^2 \Phi}{\partial t^2}-\left(\frac{m^2}{f}-\frac{q^2 A_t^2}{f^2}+q^2 A_r^2 \right)\Phi=0\,.
\ee
From this equation, we read off an effective mass squared
\beq\label{meff}
m_{{\rm eff}}^2={m^2}-\frac{q^2 A_t^2}{f}+q^2 f A_r^2.
\eeq
We will be interested in determining at what scale this becomes sufficiently negative to produce an unstable mode.

Finally, we need to impose boundary conditions. In the static case, the boundary conditions at infinity are
\be\label{eq:falloff}
\Phi = \frac{\phi_+}{r^{\Delta_+}}+ \ldots\;\;,\;\;A_t = \mu - \frac{\rho}{r}+ \ldots
\ee
where
\be
\Delta_{\pm}= \frac{3}{2}\pm \frac{1}{2}\sqrt{9+4 m^2 L^2}\,.
\ee
The fall-off for $\Phi$ corresponds to the expectation value for the operator dual to the scalar field,
\be
\langle \mc O \rangle = L^{1-2\Delta_+} \phi_+,
\ee
when there is no external source.\footnote{In the range $m_{BF}^2\le m^2 < m_{BF}^2+L^{-2}$ the mode $1/r^{\Delta_-}$ is also normalizable and we could instead consider the boundary condition $\Phi \to \phi_-/r^{\Delta_-}$.}
In order to maintain a finite current squared $J_\mu J^\mu$ at the horizon, one sets $A_t=0$ there~\cite{Hartnoll:2008vx}.

To generalize this to the time-dependent case, the first step is to replace $\mu \to \mu(t)$ and keep similar fall-offs at infinity.  In order to determine the critical temperature, we need only focus on solutions at or near $\Phi=0$; we will turn to this next.

\subsection{Estimate in a Simple Tractable Regime}\label{subsec:tractable}

We are interested in the radial scale at which the instability of the scalar field $\Phi$ kicks in; this will determine the critical temperature $T_c$ for the superconducting transition in the dual field theory.  In order to estimate
this, we can simply study the unstable solutions at $\Phi=0$ and determine at what radial scale $m_{{\rm eff}}^2$ goes sufficiently negative.  Setting $\Phi=0$ in (\ref{maxwelleom}), (\ref{maxwelleom2}), and (\ref{phieom}), the solution simplifies to
\beq\label{Fsoln}
F_{tr}=-\frac{r_+\mu_0}{r^2}\,,
\eeq
where our choice of notation will become clear momentarily.
Now let us impose that
\beq\label{Atmu}
A_t\to \mu(t)-\frac{\rho(t)}{r}
\eeq
as we approach the boundary $r\to\infty$.

Plugging (\ref{Atmu}) into (\ref{Fsoln}) yields
\beq
F_{tr}=\partial_t A_r-\partial_r A_t=\partial_tA_r-\frac{\rho(t)}{r^2}\equiv -\frac{r_+ \mu_0}{r^2},
\eeq
valid near the boundary.  The boundary condition for $A_t$ near the horizon will determine the relation between the chemical potential and charge density, leading to $\rho(t)=r_+\mu(t)$.  Anticipating this, we obtain
\beq\label{Arsol}
A_r = \frac{r_+}{r^2}\int dt'(\mu(t')-\mu_0)
\eeq
where we set the integration constant to zero so that on average $\langle A_r \rangle_t=0$, as in the static case. For a sinusoidally varying chemical potential
\be
\mu(t)=\mu_0-\delta\mu \sin(\omega t)
\ee
we have
\beq\label{Arsoln}
A_r=\frac{r_+}{r^2}\frac{\delta \mu}{\omega}\cos(\omega t)\,.
\eeq
Notice that the amplitude of $A_r$ is controlled by $\delta \mu/\omega$.

In fact, (\ref{Atmu}) and (\ref{Arsol}) solve the equations of motion everywhere in the bulk, not just near the boundary.  The boundary conditions near the horizon determine the relation between $\mu(t)$ and $\rho(t)$.   Let us impose that $A_t\to 0$ at the horizon (so that $J_\mu J^\mu$ is finite there, where $J_\mu$ is the bulk gauge current~\cite{Hartnoll:2008vx}).  This is satisfied if we set $\mu(t)=\rho(t)/r_+$, i.e.
\beq\label{Atsoln}
A_t=\mu(t)\left(1-\frac{r_+}{r}\right)\,.
\eeq

Now in the near-boundary regime $r\gg L$, the emblackening factor $f(r)\to r^2/L^2$.  From the
expression (\ref{meff}) for $m_{{\rm eff}}^2$, the charge density will enhance the instability of $\Phi$ provided that
\be
A_t^2 \, \frac{L^4}{r^4} > A_r^2\,.
\ee
Because of the factor of $\omega^{-1}$ in (\ref{Arsoln}), for sufficiently large $\omega$ this will be satisfied, and the effect is enhanced by increasing the time dependence (increasing $\omega$).

Note that $m_{{\rm eff}}^2 \to m^2$ both near the boundary and at the horizon. An instability will appear if $A_t$ can make the effective mass squared sufficiently negative over a large enough range of radial scales.  In particular, if we begin with a bare $m^2$ which is right at the Breitenlohner-Freedman bound, the net negative contributions to $m_{\rm eff}^2$ of of the time dependence must induce a condensate.

\begin{figure}[ht!]
\centering
\includegraphics[width=4in]{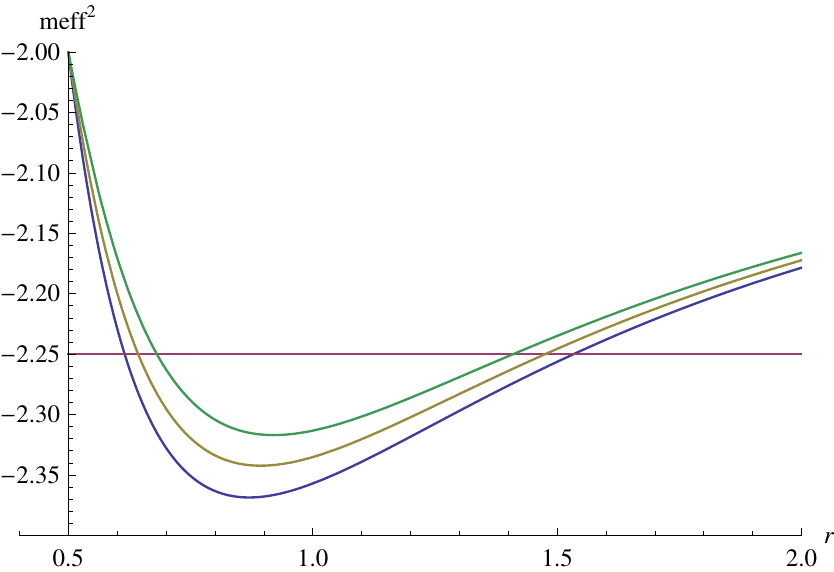}
\caption{\small{Effective mass $m_{{\rm eff}}^2(r)$ for different values of the frequency $\omega$. In the plot, $m^2L^2=-2$, $r_+/L=0.5$, and for reference the BF bound $m_{BF}^2L^2=-9/4$ is shown. Increasing the frequency makes $m_{{\rm eff}}^2$ more negative over a larger range of radial coordinate.}}
\label{fig:figure1}
\end{figure}

In \S \ref{subsec:Tc} and \S \ref{sec:condensed} we will further analyze this, and solve the equation of motion for $\Phi$ numerically near $T_c$.  Its basic features can be understood qualitatively. The instability will set in at some $r_c$ which is greater than its value in the absence of the oscillating charge density. This increases $T_c$ since it increases the radial position that needs to be cloaked by the horizon if we wish to heat the system up above the superconducting transition. These effects are illustrated in Figure \ref{fig:figure1}, where $m_{{\rm eff}}^2(r)$ is plotted for different values of the frequency.
Moreover, in the case $\mu_0=0$ we see that the time-averaged chemical potential is zero, and the superconductivity follows entirely from the oscillating contribution to the time average $\langle g^{tt}A_t^2\rangle_t$.

\subsection{Determination of $T_c$}\label{subsec:Tc}

For the purpose of finding the dependence of $T_c$ on $\delta \mu$ and $\omega$ it is sufficient to restrict to $T \approx T_c$. In this approximation, $\Phi$ should be everywhere small and its backreaction on the gauge fields (which is quadratic in $\Phi$) may be ignored. $A_r$ and $A_t$ are then given by (\ref{Arsol}) and (\ref{Atsoln}).

Furthermore, we focus on the large frequency regime
, and approximate the time-dependent equation of motion for $\Phi$ by its time average. This reduces to the ordinary linear differential equation
\be\label{phieom2}
-\frac{d}{dr}\left(r^2 f \frac{d\Phi}{dr} \right)+r^2m_{\rm eff}^2\,\Phi=0\,,
\ee
where the effective mass (\ref{meff}) becomes
\be\label{meff2}
m_{\rm eff}^2 =m^2-\frac{q^2}{f} \left(\mu_0^2 + \frac{1}{2}\delta \mu^2 \right)\left(1- \frac{r_+}{r} \right)^2+\frac{q^2}{2} \frac{\delta \mu^2}{\omega^2}\frac{r_+^2\,f}{r^4}\,.
\ee
As we noted before, our mechanism for increasing $T_c$ relies on the fact that $m_{\rm eff}^2$ is determined by the time averages $\langle g^{rr} A_r^2 \rangle_t$ and $\langle g^{tt} A_t^2 \rangle_t$, which are nonzero even if $\langle A_\mu \rangle_t=0$. We also find it convenient to set $q=1$ by rescaling $\Phi$ and $A_\mu$.

For sufficiently low $T/\mu$, we expect that (\ref{phieom2}) admits normalizable solutions. Since we are interested in the onset of the instability, we look for a marginally stable mode -- namely a solution $\Phi(r)$ which is normalizable and everywhere small.\footnote{In the static case this was studied by~\cite{Gubser:2008px}.} This also guarantees that our approximation of neglecting backreaction is consistent. The boundary conditions then read
\be\label{eq:regular}
\Phi(r \to \infty) \to \frac{\phi_+}{r^{\Delta_+}}\;,\;|\Phi(r\to r_+)|<\infty\,.
\ee
Notice that the regularity condition at the horizon is different from normalizability; this condition will play an important role shortly.

The existence of marginally stable solutions can be reduced to a quantum mechanics problem. Redefining
\be
\Phi(r) = \frac{1}{r f(r)^{1/2}} \hat \Phi(r)
\ee
sets to zero the friction term, and (\ref{phieom2}) becomes
\be\label{eq:schr1}
- \frac{d^2 \hat \Phi}{dr^2}+ V(r) \hat \Phi=0\,.
\ee
Here we have defined
\be\label{eq:schr2}
V(r) \equiv \frac{m_{\rm eff}^2}{f}+ \frac{1}{2} \frac{f''}{f}+ \frac{f'}{rf}-\frac{1}{4} \left(\frac{f'}{f} \right)^2\,,
\ee
and primes denote derivatives with respect to $r$.
Therefore, we are looking for bound states of the Schr\"odinger equation in the central potential $V(r)$, with energy $E=0$.  Note, however, that the normalization condition on $\hat\Phi$ is distinct from the usual one for the wavefunction in quantum mechanics.  (It is possible to further redefine variables to obtain the standard normalization condition, but we do not do so here for brevity.)

The strength of the negative term in the potential is controlled by the dimensionless parameter
\be\label{eq:lambda}
\lambda = \frac{\mu_0^2 + \delta \mu^2/2}{r_+^2/L^4}\,.
\ee
As we increase $\lambda$ we expect to find a critical value $\lambda_c$ at which the first bound state appears with $E=0$. This configuration corresponds to the onset of the superconducting transition and determines $T_c$:
\be
T_c=\frac{3}{4\pi} \left(\frac{\mu_0^2 + \delta \mu^2/2}{\lambda_c} \right)^{1/2},
\ee
where we used (\ref{eq:T}).

The critical eigenvalue will be a function
\be
\lambda_c=h(\Delta_+, \delta \mu^2/\omega^2),
\ee
which are the remaining parameters appearing in (\ref{eq:schr2}).\footnote{Naively one might expect the behavior of (\ref{eq:schr1}) to depend on four dimensionless parameters: $\lambda$, $\delta\mu^2/\omega^2$, $\Delta_+$, and $r_+/L$. However there is an additional scaling symmetry, under which $r$ and $r_+$ have charge 1, and everything else is neutral, ruling out $r_+/L$. This corresponds to the fact that on the dual field theory side $T_c$ should not depend on $L$.}

\begin{figure}[ht!]
\centering
\includegraphics[width=3.5in]{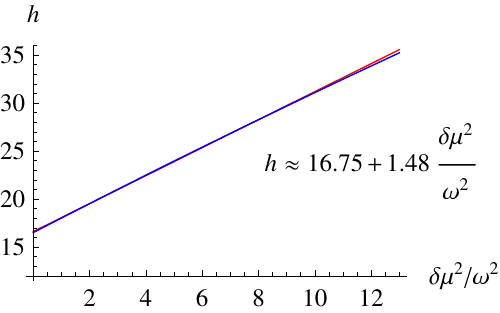}
\caption{\small{Behavior of $h$ as a function of $\delta \mu^2/\omega^2$, for $\Delta_+=2$. The blue line shows the numerical result, and the red line is the best fit, whose equation is also shown in the plot.}}
\label{fig:figure2}
\end{figure}

We performed a numerical analysis of the first marginally stable solution and the function $h$. The numerical routine imposes the boundary conditions
\be\label{eq:regular2}
\Phi(r \to \infty) \to \frac{\phi_+}{r^{\Delta_+}}\;,\;\Phi'(r_+) = \frac{1}{3} \Delta_+(\Delta_+-3) \frac{\Phi(r_+)}{r_+}\,,
\ee
and then scans over increasing values of $\lambda$ until (\ref{phieom2}) admits the first nontrivial solution.\footnote{This is similar to the static case calculations in~\cite{Hartnoll:2008vx}. We thank S. Hartnoll for discussions on this point.} We find that $h$ is well approximated by a linear function in $\delta \mu^2/\omega^2$, so we arrive at the expression for the critical temperature
\be\label{eq:Tc}
T_c\approx\frac{3}{4\pi}\left( \frac{\mu_0^2+\frac{1}{2}\delta \mu^2}{\gamma_0+\gamma_1 \frac{\delta \mu^2}{\omega^2}}\right)^{1/2}\,.
\ee
Here $\gamma_0$ and $\gamma_1$ are functions that increase monotonically with $\Delta_+$; $\gamma_0$ has been analyzed in~\cite{Horowitz:2008bn}. An example of this behavior and the fit are shown in Figure \ref{fig:figure2}, for the case $\Delta_+=2$.

\vskip 2mm

The formula (\ref{eq:Tc}) for $T_c$ reveals some interesting properties of our system. First, in both the static and time-dependent cases $T_c$ decreases as we increase $\Delta_+$. The reason for this is that increasing $\Delta_+$ increases the scalar mass, which disfavors an instability. As an example, for a fermion bilinear $\mc O = \psi \psi$, the operator dimension is $\Delta_+=2$ (at weak coupling and, with suitable supersymmetric protection, also at strong 'tHooft coupling).  In this case, the critical temperature becomes
\be
T_c\approx\frac{3}{4\pi} \left( \frac{\mu_0^2+\frac{1}{2}\delta \mu^2}{16.75+1.48 \frac{\delta \mu^2}{\omega^2}}\right)^{1/2}\,.
\ee

We also see that a slowly-varying chemical potential degrades the superconducting properties of the system, by lowering the value of $T_c$; however, above a certain critical value $\omega>\omega_c$, the time-dependent source enhances superconductivity. Demanding $T_c>T_c^{(0)}$, where $T_c^{(0)}=\frac{3}{4\pi}\mu_0/\sqrt{\gamma_0}$ is the transition temperature in the static case, we obtain
\be\label{eq:omegac}
\omega_c=\sqrt{\frac{2\gamma_1}{\gamma_0}}\mu_0=\frac{4\pi}{3}\sqrt{2\gamma_1}T_c^{(0)}\,.
\ee

A critical frequency $\omega_c$ also arises in BCS superconductors, as first recognized by Eliashberg~\cite{Eliashberg}\ who found
\beq
\omega_c^2\sim \frac{\Delta}{\tau_0}
\eeq
where $\tau_0$ is the relaxation time of the fermions, and $\Delta$ is the gap.  Roughly speaking, one must excite the system at a rate fast enough to compete with the dissipation of the quasiparticles.

\section{Analysis of the Condensed Configuration}\label{sec:condensed}

Having determined that the instability is enhanced by virtue of time dependence, we now consider the solution where the scalar field condenses, for temperatures $T< T_c$. We first describe the basic features of the condensed solution and analyze the order parameter $\langle \mc O \rangle \propto \phi_+$. Next, we check that our classical time-averaged analysis provides a good approximation to the behavior of our system.

\subsection{Description of the Condensed Solution}

So far we have determined that $T_c$ has increased according to the classical time-averaged solution, but have not exhibited the condensed solution itself for $T< T_c$.  In the full solution, the gauge fields are deformed away from (\ref{Atsoln}) and (\ref{Arsol}), responding to the nonzero $\Phi^2$ in the last term of each of equations (\ref{maxwelleom}) and (\ref{maxwelleom2}).  These effects have to be taken into account in order to determine the condensed configuration. Indeed, the backreaction $\delta A_\mu \propto \Phi^2$ generates nonlinear terms in $\Phi$. Balancing them against the tachyonic mass produces a nontrivial profile for $\Phi$ that we now study.

In the large frequency regime the deformation of $A_r$ is suppressed by additional powers of $1/\omega$ and may be neglected. The deformation of $A_t$ is sourced by the term $\Phi^2 A_t$ in the Maxwell equation.
This is a rapidly oscillating term, controlled by $\mu(t)$, but in order to capture the backreaction on the time-averaged solution for $\Phi$, we must include corrections to the rapidly oscillating solution for $A_t$.
As in our discussion above of the unperturbed solution, the interaction of the gauge field with $\Phi$ is proportional to $A_t^2$, which does not vanish on time-average; therefore, we must solve the Maxwell equation for the back reaction including the fast oscillations, and then solve the $\Phi$ equation in a time-averaged approximation.

Let us now consider the condensed solution. It is important to analyze the behavior of the order parameter $\phi_+$ near $T \approx T_c$ first, as this characterizes the phase transition in the dual superconductor. In this limit, the backreaction on the gauge field is small, so we only need to keep the leading effect. To work this out, we vary a little bit away from $T_c$, and the resulting scalar field must be small: $\Phi\approx\epsilon\Phi_0$, where $\Phi_0$ is the normalized solution to equation (\ref{phieom2}) in the non-backreacted case. The leading backreaction to $A_t$ can be determined from its equation of motion as
\begin{equation}
A_t=[\mu_0-\delta\mu\sin(\omega t)]\left\{\left(1-\frac{r_+}{r}\right)+\epsilon^2\left[h(r)-h(r_+)\frac{r_+}{r}\right]\right\}
\end{equation}
where $h(r)$ is defined as
\begin{equation}
h(r)=\int_r^\infty\frac{dr'}{r'^2}\int_{r'}^\infty dr''\frac{2r''^2}{f(r'')}\left(1-\frac{r_+}{r''}\right)\Phi_0(r'')^2
\end{equation}
and the $h(r_+)r_+/r$ term makes sure that $A_t$ vanishes at the horizon. Here the term of order $\epsilon^2$ is the deformation of $A_t$ due to the backreaction. Let us note that this deformation is negative,\footnote{This can either be seen numerically, or shown by changing coordinates from $r$ to $z=r_+/r$. Then $h$ is a concave function in $z$ vanishing at $z=0$. The deformation is $h(z)-h(z=1)z$ and therefore non-positive for any $z$ or $r$.} therefore making a positive contribution to $m_{\textnormal{eff}}^2$ in (\ref{phieom2}) that is of order $\epsilon^2$. This must be compensated by varying $T$ away from $T_c$ by an appropriate (negative) amount, which changes (\ref{phieom2}) by terms of order $T-T_c$. Therefore, $\epsilon$ must be proportional to $(T-T_c)^{1/2}$. Putting back the dimensional parameters, the value of the order parameter in the large frequency limit becomes
\be\label{eq:phip}
\langle \mc O \rangle =\frac{\phi_+}{L^{2\Delta_+-1}} \approx c_{\Delta_+}  T_c^{\Delta_+}\left(1- T/T_c \right)^{1/2}
\ee
with $c_{\Delta_+}$ a monotonically increasing function of $\Delta_+$.
The main features of this result are the overall factor $T^{\Delta_+}$, which reflects the dimension of the operator $\mc O$ dual to $\Phi$, and the dependence $(1-T/T_c)^{1/2}$. This is the expected critical exponent from a mean-field description of the superconducting transition. This expression for $\mc O$ is formally the same as in~\cite{Hartnoll:2008vx}, but with $T_c$ now enhanced by time-dependence, Eq.~(\ref{eq:Tc}).

On the gravity side we can go beyond the LG approximation and study the system at $T \ll T_c$, where it is found that the order parameter has the limiting behavior $\langle \mc O \rangle \propto T_c^{\Delta_+}$. We refer the reader to~\cite{Hartnoll:2008vx} for more details.

\subsection{Classical and Quantum Corrections to the Time-Averaged Approximation}

We find that the time-averaged approximation is a valid approximation at least for large frequency $\omega$.  To check this we first consider the effects of relaxing the time averaging on the classical solution.  Then we estimate quantum particle production and decay.

\subsubsection{Classical Solutions}

Let us consider the time-dependent equation of motion (\ref{phieom}) for the scalar $\Phi$ in the background gauge field (\ref{Arsoln}) and (\ref{Atsoln}). As argued earlier, this is the sufficient for determining $T_c$, because at $T\approx T_c$ the backreaction of $\Phi$ on the gauge field is negligible. We may expand the full solution in Fourier series (anticipating that the system will only source modes at multiples of the frequency $\omega$)
\begin{equation}
\Phi(r,t)=\phi_0(r)+\phi_1(r)\sin(\omega t)+\cdots
\end{equation}
The time-averaged approximation allows us to neglect $\phi_1$ (and all higher Fourier modes). Let us work out the corrections if we had included $\phi_1$. At this order of Fourier expansion, $\phi_1$ satisfies the following ordinary differential equation
\begin{equation}\label{phi1eom}
-\frac{1}{r^2}\frac{d}{dr}\left(r^2 f \frac{d\phi_1}{dr} \right)+m_{\textnormal{eff}}^2\phi_1-\frac{\omega^2}{f}\phi_1+\frac{2q^2\mu_0\delta\mu}{f}\left(1- \frac{r_+}{r} \right)^2\phi_0=0,
\end{equation}
where $m_{\textnormal{eff}}^2$ is defined by (\ref{meff2}). This is the same equation as (\ref{phieom2}) save for the addition of the last two terms. The last term is a source for $\phi_1$; without it we would have the static case with vanishing $\phi_1$. Therefore, we expect $\phi_1$ to be linear in this source term (at least when it is small). At large frequency $\omega$ the last two terms compete with each other, giving
\begin{equation}\label{phi1sol}
\phi_1\approx\frac{2q^2\mu_0\delta\mu}{\omega^2}\left(1- \frac{r_+}{r} \right)^2\phi_0<\frac{2q^2\mu_0\delta\mu}{\omega^2}\phi_0\,.
\end{equation}
We can also work out the equation satisfied by the zero mode $\phi_0$:
\begin{equation}\label{phi0eom}
-\frac{1}{r^2}\frac{d}{dr}\left(r^2 f \frac{d\phi_0}{dr} \right)+m_{\textnormal{eff}}^2\phi_0+\frac{q^2\mu_0\delta\mu}{f}\left(1- \frac{r_+}{r} \right)^2\phi_1=0,
\end{equation}
where the last term is a source. This term is the leading correction to the time-averaged approximation. We can estimate its effect by taking its ratio to a typical term in the equation, e.g. the $\mu_0^2+\frac12\delta\mu^2$ term inside $m_{\textnormal{eff}}^2$. According to (\ref{phi1sol}) this ratio is
\begin{equation}
\frac{\mu_0\delta\mu\phi_1}{(\mu_0^2+\frac12\delta\mu^2)\phi_0}<\frac{2q^2\mu_0^2\delta\mu^2}{(\mu_0^2+\frac12\delta\mu^2)\omega^2}~.
\end{equation}
So as long as the frequency $\omega$ is much larger than either $q\mu_0$ or $q\delta\mu$, the correction to the time-averaged approximation is small.

\subsubsection{Stability and (Un)particle Production}

So far we have considered the behavior of the classical solution in the presence of a time dependent chemical potential.  Time-dependent perturbations of a field theory generically lead to additional quantum effects such as particle production (or in the case of conformal field theory, `unparticle' production) as well.
In this section we would like to estimate these additional effects of the time-dependent chemical potential $\delta\mu(t)$ in our system, expanded around the classical solution $\Phi^{(0)}(r,t), A_\mu^{(0)}$ for the condensed scalar $\Phi$ and the gauge fields $A_\mu$.
To estimate particle production on the gravity side we must study the perturbations without time-averaging the chemical potential.

Let us work in a locally flat frame with coordinates $T\sim t r/L, \vec X\sim \vec x r/L,R\sim L \,\log(r/L)$.
Expanding about the classical solution $\Phi^{(0)},A_\mu^{(0)}$,  we expect the scalar field perturbation $\delta\Phi$ to be stable in the time averaged case as in the case of static holographic superconductors.
Consider a toy model for our problem encoding this feature, in which the Fourier modes $\delta\Phi_k$ locally satisfy an equation of motion of the form
\beq\label{delphi}
\frac{d^2\delta\Phi}{dT^2} + \Omega_K^2(t)\delta\Phi = 0, ~~~~~{\rm with}~~~ \Omega_K^2(t)= \frac{L^2}{r^2}\delta\mu^2(\eta_0/2-\cos^2(\omega L T/r))+\vec K^2.
\eeq
Here we take $\eta_0\ge 1$ to model the fact that the lightest perturbation is stable in the time-averaged case.
We read off the coefficient $L^2\delta\mu^2/r^2$ of the perturbation from the action expanded in our locally flat frame.

The first question that this raises is whether
a tachyonic mode develops during the times when $\Omega_k^2(t)<0$ (which occurs for sufficiently small proper momentum $K$). 
Such a tachyon grows at most exponentially as $e^{\delta \mu L T/\sqrt{2}r}=e^{\delta\mu t/\sqrt{2}}$.  The timescale $\sim 1/\delta \mu$ for this to grow appreciably is much slower than the period $\sim\omega^{-1}$ over which the mode remains tachyonic, so the perturbation does not grow exponentially (see e.g. \cite{Dvali:2003vv}\ for a recent discussion of this effect).

The next question is whether there is significant particle production, and if so what its consequences are.  In particular, the greatest production occurs near the resonance frequency.
At the Gaussian level for the perturbations $\delta\Phi$ of the charged scalar, to calculate particle production one would read off the Bogoliubov coefficients from the solutions to (\ref{delphi}).  At leading order in the perturbation, the pair production can be computed using ordinary quantum mechanics by analyzing the two-state system consisting of the ground state and the two-particle state.
For energy levels which are separated by an energy difference from the vacuum of near the resonance frequency,
such a two-state system oscillates between the vacuum and the excited state \cite{LL}, with a proper frequency of order $\omega_{slow}\equiv\delta\mu^2 L/\omega r$  (see e.g. \cite{Flauger:2009ab}\ for a recent discussion of this effect applied to particle production).  So for sufficiently short timescales this is a small effect and the probability amplitude is of order the nontrivial Bogoliubov coefficient whose square gives the number density of produced particles

The resulting proper energy density in produced $\delta\Phi$ particles near resonance is of order
\beq
\rho_{\delta\Phi}\sim \omega_p^4 (\omega_{slow} T)^2 \sim \omega_p^2 \langle \delta\Phi^2\rangle
\eeq
where we defined the proper frequency $\omega_p=\omega L/r$.  Given this result for $\langle\delta\Phi^2\rangle$,
we can plug it into (\ref{maxwelleom}) to determine on what timescale $t_{backreaction}$ the produced particles source a deformation of $A_t$ which competes with its original magnitude $\sim \delta\mu$.  This timescale
depends on the radial position in the bulk at which we analyze it.  Near $r_c\sim T_c L^2$, it turns out to be
\beq\label{tbr}
t_{backreaction}\sim T_c/\delta\mu^2\sim 1/\delta\mu.
\eeq

Although parametrically longer than $\omega^{-1}$ in our setup, this timescale is very short if we think of our system as a toy model for a real material.

However, the quartic coupling $\delta\Phi^2 A_\mu^2$ leads to decay of $\delta\Phi$ modes into gauge bosons.  This rate $\Gamma$ is of order $\omega_p=\omega L/r$, and is much greater than $\omega_{slow}$ above  precisely for $\omega^2\gg \delta\mu^2$, our regime of interest.
So in our regime of parametrically large $\omega^2/\delta\mu^2$ the produced $\delta\Phi$ particles decay long before their number builds up to the point where $\delta\Phi$ backreacts on the solution.  The decay products consist of a superposition of high-frequency electromagnetic waves.  These themselves can annihilate back into $\Phi$ particles and also can mix with other sectors (for example, neutral moduli fields, other gauge bosons, and other charged fields).  The net effect of this on $m_{eff}^2$ and on the scalar condensate depends on the details of these sectors, but it is clear that with sufficient mixing to modes neutral under the gauge symmetry corresponding to $A_\mu$, this backreaction can be small.

Note that the energy density of produced particles can remain well below the gravity-side Planck density, and also well below the energy density $M_{Planck}^2/L^2$ which would source significant corrections to the background geometry.  This means that the system is not substantially heated by the process until the energy builds up to an extent that the probe approximation breaks down, something we can formally push parametrically as far away as we wish.  In real systems, it has been argued \cite{Eliashberg}\cite{SF}\ that a steady state can be reached, balancing dissipation and time-dependent excitation.  It would be interesting to explore this upon relaxing the probe approximation in the strongly-coupled systems we are studying.

\section{Discussion}

It is interesting to compare and contrast the stimulated superconductivity in the holographic theories of \S\S 3 and 4 to the theory reviewed in \S2.  In the former, the constituents of the scalar condensate, to the extent that that notion applies, are strongly bound.  In the latter, the kinematics of the Fermi surface plays a key role.  In both cases, the effect is sensitive to the square of the time-dependent perturbation.  In the holographic case, this survives time-averaging and contributes to $m_{\rm eff}^2$.  In the BCS case,
the condensate determined by equations (\ref{gap}) and (\ref{shiftn}) is linear in $\delta n$, and $\delta n$ is determined by the probability of up-scattering away the quasiparticles occupying states near the Fermi surface at finite temperature.  This is the square of the scattering amplitude induced proportional to the time dependent term in the Hamiltonian. Of course, our system is on the face of it very far from BCS theory, manifesting only the physics of a bosonic superfluid order
parameter.

In the present analysis, the only limitation on the size of the time-dependent source is the probe approximation.
This approximation includes arbitrary interactions of the bulk matter and bulk gauge fields, which affect each other significantly but do not backreact on the background geometry (which is determined by an ambient strongly coupled large-N sector).  This is a rather different regime of control than that obtained in the previous works such as \cite{Eliashberg}\cite{SF}, which were limited to a treatment perturbative in $\delta n$ (\ref{shiftn}).  One question raised in those works was what happens beyond that approximation;  in the holographic examples, the beneficial effects of time dependence get stronger in this regime, as long as we simultaneously increase $\omega$, keeping $\omega/\delta\mu$ large.

Given this, it would also be very interesting to go beyond the probe approximation, and study the effects of our time-dependent source including gravitational backreaction. The static superconductor beyond the probe limit was analyzed in \cite{Hartnoll:2008kx}, where it was found that the superconducting instability persist also in the backreacted solution. A first step in this direction would be to define a consistent time-averaged approximation including deformations of the metric. Such an analysis would extend our mechanism to systems that do not admit a probe limit, and may reveal new constraints on the increase on $T_c$ by time-dependent effects.

So far we have focused on a sinusoidally oscillating chemical potential.  Perhaps it is more realistic to consider irradiating the system with electromagnetic radiation.  This can be analyzed similarly; it introduces various contributions to $m_{eff}^2$ through $\langle A_t^2\rangle$ and $\langle A_{spatial}^2\rangle$.  

\section*{Acknowledgments}
We are grateful to A.~Adams, S.~Das, S.~Hartnoll, G.~Horowitz, X.~-L.~Qi and S.~Yaida for very useful comments. This research was supported in part by the National Science Foundation under grant PHY05-51164, by NSF grant PHY-0244728, and by the DOE under contract DE-AC03-76SF00515.


\bibliographystyle{JHEP}
\renewcommand{\refname}{Bibliography}
\addcontentsline{toc}{section}{Bibliography}
\providecommand{\href}[2]{#2}\begingroup\raggedright

\end{document}